\documentclass[apsprb,twocolumn,superscriptaddress,amsmath,amssymb,showpacs]{revtex4-1}
\usepackage{graphicx}% Include figure files
\usepackage{bm}% bold math
\usepackage{txfonts}

\begin{document}
%\title{Investigation of transverse nuclear field formation in single self-assembled quantum dots: \\
%role of nuclear quadrupole effects}
\title{Investigation of in-plane nuclear field formation in single self-assembled quantum dots}

\author{S.\ Yamamoto}
	\affiliation{Division of Applied Physics, Hokkaido University, N13 W8, Kitaku, Sapporo 060-8628, Japan}
\author{R.\ Matsusaki}
	\affiliation{Division of Applied Physics, Hokkaido University, N13 W8, Kitaku, Sapporo 060-8628, Japan}
\author{R.\ Kaji}
%\email{r-kaji@eng.hokudai.ac.jp}
	\affiliation{Division of Applied Physics, Hokkaido University, N13 W8, Kitaku, Sapporo 060-8628, Japan}	
\author{S.\ Adachi}
\email{adachi-s@eng.hokudai.ac.jp}
	\affiliation{Division of Applied Physics, Hokkaido University, N13 W8, Kitaku, Sapporo 060-8628, Japan}

\date{\today}% It is always \today, today,

%%%%%%%%%%%%%%%%%%%%%%%%%%%%%%%%%%%%%%
\begin{abstract}
We studied the formation mechanism of the in-plane nuclear field in single self-assembled In$_{0.75}$Al$_{0.25}$As/Al$_{0.3}$Ga$_{0.7}$As quantum dots. The Hanle curves with an anomalously large width and hysteretic behavior at the critical transverse magnetic field were observed in many single quantum dots grown in the same QD sample. In order to explain the anomalies in the Hanle curve indicating the formation of a large nuclear field perpendicular to the photo-injected electron spin polarization, we propose a new model based on the current phenomenological model for dynamic nuclear spin polarization. The model includes the effects of the nuclear quadrupole interaction and the sign inversion between in-plane and out-of-plane g-factors, and the model calculations reproduce successfully the characteristics of the observed anomalies in the Hanle curves.
% Based on the current phenomenological model for dynamic nuclear spin polarization, we proposed a new model including the nuclear quadrupole interaction and the sign inversion between in-plane and out-of-plane g-factors in order to explore the formation mechanism of a considerable nuclear field perpendicular to the photo-injected electron spin polarization. The model calculations reproduced successfully the characteristics of the observed anomalies in the Hanle curves. 
\end{abstract}
%%%%%%%%%%%%%%%%%%%%%%%%%%%%%%%%%%%%%%
\pacs{73.21.La, 78.67.Hc, 71.35.Pq, 71.70.Jp}% PACS, the Physics and Astronomy
%\keywords{nuclear spin, Overhauser field, quantum dots, hyperfine interaction}
%Use showkeys class option if keyword
                              %display desiredc
\maketitle

\section{Introduction}\label{intro}

The study of nuclear spin physics in semiconductor quantum dots (QDs) is an active research field currently. 
This is because the role of hyperfine interaction (HFI), which is the magnetic interaction between a localized electron and the lattice nuclei, is drastically enhanced in QD structures compared with those in bulks and quantum wells due to a strong localization of the electron wave function~\cite{OptOrientation,SpinPhysics,Urbaszek13}. Since it is possible to transfer the angular momentum from light onto nuclei via electron spin, a macroscopic nuclear spin polarization (NSP) which is orders of magnitude larger than the value in thermal equilibrium can be generated actually at cryogenic  temperatures, and in turn, the resultant nuclear field (Overhauser field, $\bm{B}_{\rm n}$) up to a few Teslas affects the electron spin dynamics significantly~\cite{Gammon01,Yokoi05,Eble06,Braun06,Tartakovskii07,Maletinsky07,Kaji08}. 
Because the lattice nuclei act as a reservoir for an optically or electrically injected electron spin, the \textit{engineering} of nuclear spins such as the optical manipulation of the NSP not only leads to the potential applications but also opens up a new spin physics.

The dynamics of NSP is determined by the environment which the nuclei are exposed to, such as presences of an external magnetic field and/or an unpaired electron in a QD, and dipole-dipole interaction among the neighbor nuclei. In particular, nuclear quadrupole interaction (QI), which originates from the coupling of a nuclear spin with $I > 1/2$ to the electric field gradients (EFG)~\cite{Slichter96}, has received a lot of attention recently. Since the lattice strain is used as a driving force for the spontaneous formation process of self-assembled QDs (SA-QDs), the residual strain and the resultant large EFG arise in these QDs. Therefore, the impact of QI is expected to gain considerably and to play key roles for various novel phenomena observed in SA-QDs~\cite{Latta09,Xu09,Hogele12,Krebs10}. The QI yields the non-equivalent energy splitting depending on the value $|I_{z'}|$, where $z'$-axis is the quantization axis determined by EFG and is found to be usually close to the sample growth axis ($z$-axis) in the SA-QDs~\cite{Chekhovich12,Chekhovich15}. Then, QI can be treated as an effective field affecting the nuclear spins, the quadrupolar field, and it stabilizes the NSP in $z'$-axis as reported in ensembles of SA-InP/InGaP QDs~\cite{Dzhioev07} and single SA-InAs/GaAs QDs~\cite{Maletinsky08}.

Recently, the formation of in-plane nuclear field was reported in single SA-InAs/GaAs QDs under quasiresonant excitation by Krebs et al.~\cite{Krebs10} and nonresonant excitation by Nilsson et al.~\cite{Nilsson13}, and it seemed to be related also to QI. The in-plane nuclear field was detected by observing the electron spin depolarization curve in Voigt configuration (i.e. Hanle curve).  In their pioneering works, the Hanle curve was distorted drastically from a normal Lorentzian shape in the following respects: a $\sim$20 times larger width than the one expected from the electron spin lifetime, and the abrupt change in the degree of circular polarization (DCP), and thus, the anomalous Hanle curve has a shape like a circus tent. 
In addition, the fact that such anomalies in Hanle curve have not been observed in single droplet GaAs QDs~\cite{Sallen14} which is free from internal strain suggests that the QI contributes significantly to an anomalous Hanle curve observed in SA-QDs. However, the origin of the anomalies in the Hanle curves has not been revealed entirely. The knowledge of the in-plane nuclear field formation may lead directly to an optical control of the nuclear field direction, and therefore, it is very important. 

In this work, we study the formation mechanism of the in-plane nuclear field via the Hanle effect measurements and model calculations. The anomalously distorted Hanle curves similar to Ref.~\onlinecite{Krebs10} are observed in single InAlAs QDs, and we show that the anomalies of the Hanle curves can be reproduced qualitatively by a proposed model including the QI and the sign inversion of g-factors.

\section{QD Sample and Anomalous Hanle Curves}\label{experiment}
SA-In$_{0.75}$Al$_{0.25}$As/Al$_{0.3}$Ga$_{0.7}$As QDs grown on (100)-GaAs substrate by molecular beam epitaxy were used in this study. The QDs have a lens-shaped profile, and the typical diameter and height were evaluated to be $\sim$20 nm and $\sim$4 nm, respectively, by the atomic force microscopy measurements of a reference uncapped QD layer and the cross-section transmission electron microscope observation. %The number of nuclei in a single QD is estimated to be $N$$\sim$$3\times10^4$. 
After the fabrication of small mesa structures, the micro-photoluminescence (PL) measurements under the transverse magnetic fields ($B_{x}$) up to 1 T were carried out at 6 K. The QD sample was excited by a cw Ti:sapphire laser tuned to $\sim$730 nm, which corresponds to the transition energy to the foot of the wetting layer of the QDs. 

The polarization of an excitation beam was adjusted carefully to the circular polarization by a combination of a linear polarizer, a half wave plate, and a quarter wave plate (QWP). The circularly polarized ($\sigma^{+}$, $\sigma^{-}$) PL components were converted to the linearly polarized ones ($\pi^{x}$, $\pi^{y}$) by another QWP inserted into the detection path, and they were displaced spatially from each other by a beam displacer. Each displaced PL component was dispersed by a spectrometer and was focused on a different area of the Si-CCD detector. 
Therefore, the energy splitting between the $\sigma^{+}$ and $\sigma^{-}$ components  %Overhauser shift (OHS) 
and the DCP of the PL spectra can be acquired by a single exposure process. The details of the experimental apparatus is seen in Ref.~\onlinecite{Matsusaki17}. 
The energy resolution of our measurement system is $\le 5\ \mu\mbox{eV}$ by spectral fitting.

In this work, we focus on the PL of positive trion ($X^+$). %The polarization-resolved PL spectra in a typical single InAlAs QD are seen in Fig.~\ref{GF} (Appendix A). 
 The $X^+$ ground state consists of two holes in a spin-singlet state and one electron, and thus, the DCP of this charge state is determined only by the electron spin right before radiative recombination. Here, the DCP  is defined as $\displaystyle \rho_{\rm c} = ({I^- - I^+})/({I^- + I^+})$, where $I^{+(-)}$ represents the integrated PL intensity of $\sigma^{+(-)}$ component. The DCP is related directly to the projection of the averaged electron spin along the sample growth axis, $\langle S_z \rangle$, and the relation $\langle S_z \rangle = {\rho_{\rm c}}/{2}$ is held for $X^+$ case.

\begin{figure}[tb]
\begin{center}
\includegraphics[width=160pt]{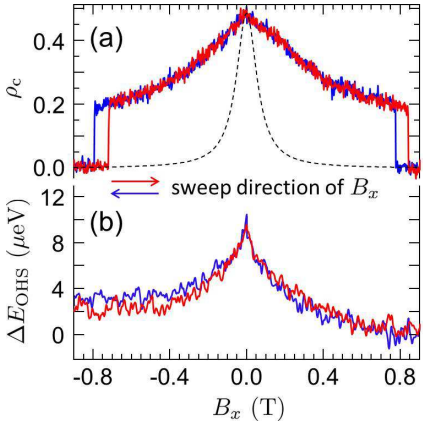}
\end{center}
\caption{(color online) (a) The anomalous Hanle curves observed in the SA-InAlAs QD. The red (blue) curve represents the DCP of $X^{+}$ PL with increasing (decreasing) $B_{x}$. The dotted curve is the normal Hanle curve expected with typical values of spin lifetime and g-factor of electron. (b) The observed Overhauser shift as a function of $B_{x}$, which is used as a measure of $|B_{{\rm n}, z}|$.}
\label{AHC}
\end{figure}

Figure~\ref{AHC}(a) is an example of the anomalous Hanle curves observed in the SA-InAlAs QD sample. 
In the figure, the red (blue) curve represents the observed DCP with increasing (decreasing) $B_{x}$ under $\sigma^-$ excitation, while the dashed curve is an expected \textit{normal} Hanle curve, which is free from the effect of $B_{\rm n}$ and has a Lorentzian shape with the full width $2B_{1/2}$($\sim$130 mT)~\cite{comment}. As clearly shown, the observed curves have quite larger widths than the expected one.
%Figure~\ref{AHC} (a) shows the observed anomalous Hanle curves in the InAlAs QD. Here, the red (blue) curve represents the result with increasing (decreasing) $B_{x}$ where the sweep rate of $B_{x}$ is 0.1 T/min.The dashed curve is the expected \textit{normal} Hanle curve, which is free from the effect of $B_{\rm n}$ and has a Lorentzian shape with the full width $2B_{1/2}$(=130 mT) determined by the electron spin lifetime $T_{\rm s}$=0.5 ns as a typical value for a single QD.The observed curves have a quite larger width than that of the expected one. 
Further, the DCP changes suddenly at the critical field $|B_x^{\rm c}|$$\sim$0.8 T, and $|B_{x}^{\rm c}|$ is different depending on the sweep direction of $B_{x}$ (i.e. a hysteretic behavior occurs). Such anomalous characters were observed in not only a specific QD but also all the QDs we measured in the same sample (not shown here). In the work by Krebs et al.~\cite{Krebs10}, the Hanle curves with the similar anomalies were observed in single SA-InAs/GaAs QDs, where the bistable behavior of a nuclear field in $x$-direction $B_{{\rm n},x}$ and a compensation of $B_x$ by $B_{{\rm n},x}$ were confirmed unambiguously.

The observed energy splittings between the $\sigma^{+}$ and $\sigma^{-}$ PL components are plotted in Fig.~\ref{AHC}(b). Since the external magnetic field is applied in $x$-$y$ plane, the energy splitting detected in ($\sigma^{+}$, $\sigma^{-}$) basis is determined  only by $z$-component of the nuclear field, $B_{{\rm n}, z}$. Thus, we term the corresponding energy splitting the Overhauser shift, $\Delta E_{\rm{OHS}}$. As seen in the figure, the maximal value of $\Delta E_{\rm{OHS}}$ of $\sim$10 $\mu$eV appears at $B_{x}$=0 T, and it corresponds to the $B_{{\rm n}, z}$ of $-$0.5 T by considering the relation $B_{{\rm n}, z}$=$\Delta E_{\rm{OHS}} / ( g_{z}^{\rm e} \mu_{\rm B} )$, where ${g}^{\rm e}_z$ is the electron g-factor in $z$-direction and $\mu_{\rm B}$ is the Bohr magneton. $\Delta E_{\rm{OHS}}$ reduces gradually with increasing $|B_{x}|$ and approaches almost zero around $|B_{x}^{\rm c}|$. Further, the observed $\Delta E_{\rm{OHS}}$ is slightly asymmetric with respect to the sign in $B_{x}$.

The set of the electron g-factor of this single QD, (${g}_{x}^{\rm e}$, ${g}_{y}^{\rm e}$, ${g}_{z}^{\rm e}$)  is described in Appendix A, and it is required for the evaluation of the nuclear field and the model calculations in the next section.

\section{Model Calculations and Discussions} \label{Model}

In this section, we propose a dynamics model of a coupled electron-nuclear spin system in order to reproduce the observed anomalous Hanle curves shown in the previous section.

The evolution of an electron spin polarization $\langle \bm{S} \rangle$ can be described by the Bloch equation:
\begin{equation}
\frac{d \langle \bm{S} \rangle}{dt} = \frac{{\bar{g}}_{\rm e}\mu_{\rm B}}{\hbar}\bm{B}_{\rm T}^{\rm {(e)}}\times \langle \bm{S} \rangle - \frac{\langle \bm{S} \rangle - \bm{S}_0}{T_{\rm s}} , \label{Bloch}
\end{equation}
where $\bar{g}_{\rm e}$ is the electron g-tensor, $\bm{B}_{\rm T}^{\rm{(e)}}$ is an effective magnetic field seen by a QD electron, and $\bm{S}_{0} =(0,0, S_{0})$ is the average electron spin polarization in the absence of $\bm{B}_{\rm T}^{\rm{(e)}}$. The first and the second terms in the right-hand side represent the Larmor precession and the electron spin relaxation with a characteristic time $T_{\rm s}$, respectively. The steady state solution of Eq.~(\ref{Bloch}) gives a Hanle curve according to the relation $\rho_{\rm c} =2 \langle S_{z} \rangle$ (see Appendix~\ref{APDXB}). Since the nuclear field $\bm{B}_{\rm n}$ ($\propto \langle \bm{I} \rangle $: NSP)  as well as the externally applied field $\bm{B}_{\rm{ext}}$ contributes to $\bm{B}_{\rm T}^{\rm{(e)}}$ ($\equiv \bm{B}_{\rm{ext}} + \bm{B}_{\rm n}$), it is essential to consider the spin dynamics of nuclei. In this work, the external field is applied in the sample growth plane, and thus, it can be written as $\bm{B}_{\rm{ext}} = (B_{x}, 0, 0)$.

As widely accepted, the contact-type HFI includes the flip-flop term between an electron and nuclear spins, i.e., $\propto ({\hat{I}_{+} \hat{S}_{-} +\hat{I}_{-} \hat{S}_{+}})$, and it allows the spin transfer from the optically-injected electron to the lattice nuclei system~\cite{OptOrientation}. The dynamics of the NSP component $\langle I_{k} \rangle$ $(k=x, y, z)$ follows the phenomenological equation~\cite{SpinPhysics,Urbaszek13}: 
\begin{equation}
\frac{d \langle I_{k} \rangle}{dt} = \frac{1}{T_{{\rm NF},k}}\left[Q (\langle S_{k} \rangle -\langle S_{k}^{\rm eq} \rangle)-\langle I_{k} \rangle \right]-\frac{1}{T_{{\rm ND},k}} \langle I_{k} \rangle, \label{eq2}
\end{equation}
where $Q=\tilde{I}(\tilde{I}+1)/[S(S+1)]$ is a numerical constant of the momentum conversion, $\langle S_{k}^{\rm eq} \rangle$ is the electron spin polarization at thermal equilibrium, and $1/T_{{\rm ND},k}$ is the relaxation rate of the NSP in $k$-direction. 
The NSP formation rate $1/T_{{\rm NF},k}$ depends on the NSP itself, and it is given by
\begin{equation}
\frac{1}{T_{{\rm NF},k}} = 2f_{\rm e}\tau_{\rm c} \left( \frac{\tilde A_k}{N \hbar} \right)^2 \left/ \left\{ 1+\left[ {g}^{\rm e}_{k} \mu_{\rm B} \left( B_{{\rm ext},k}+B_{{\rm n},k} \right) \tau_{\rm c} /\hbar \right]^2 \right\} \right. ,\label{eq3}
\end{equation}
where $f_{\rm e}$ ($0 \le f_{\rm e} \le 1$) is  a filling factor representing the occupation of a QD by an unpaired electron spin, $\tau_{\rm c}$ is a correlation time of HFI, $N$ is the number of nuclei related to the interaction, $\tilde{A}_k$ is the averaged coupling constant of HFI, and $B_{{\rm n},k} = {\tilde{A}_k I_k}/({{g}^{\rm e}_k \mu_{\rm B}})$ is the $k$-component of the nuclear field, respectively. This spin dynamics model has achieved success to explain the experimental observations about $\langle I_{z} \rangle$ and thus the behavior of $B_{{\rm n},z}$ in the previous studies~\cite{Eble06,Braun06,Tartakovskii07,Maletinsky07,Kaji08}.

\begin{figure}[t]
\begin{center}
\includegraphics[width=240pt]{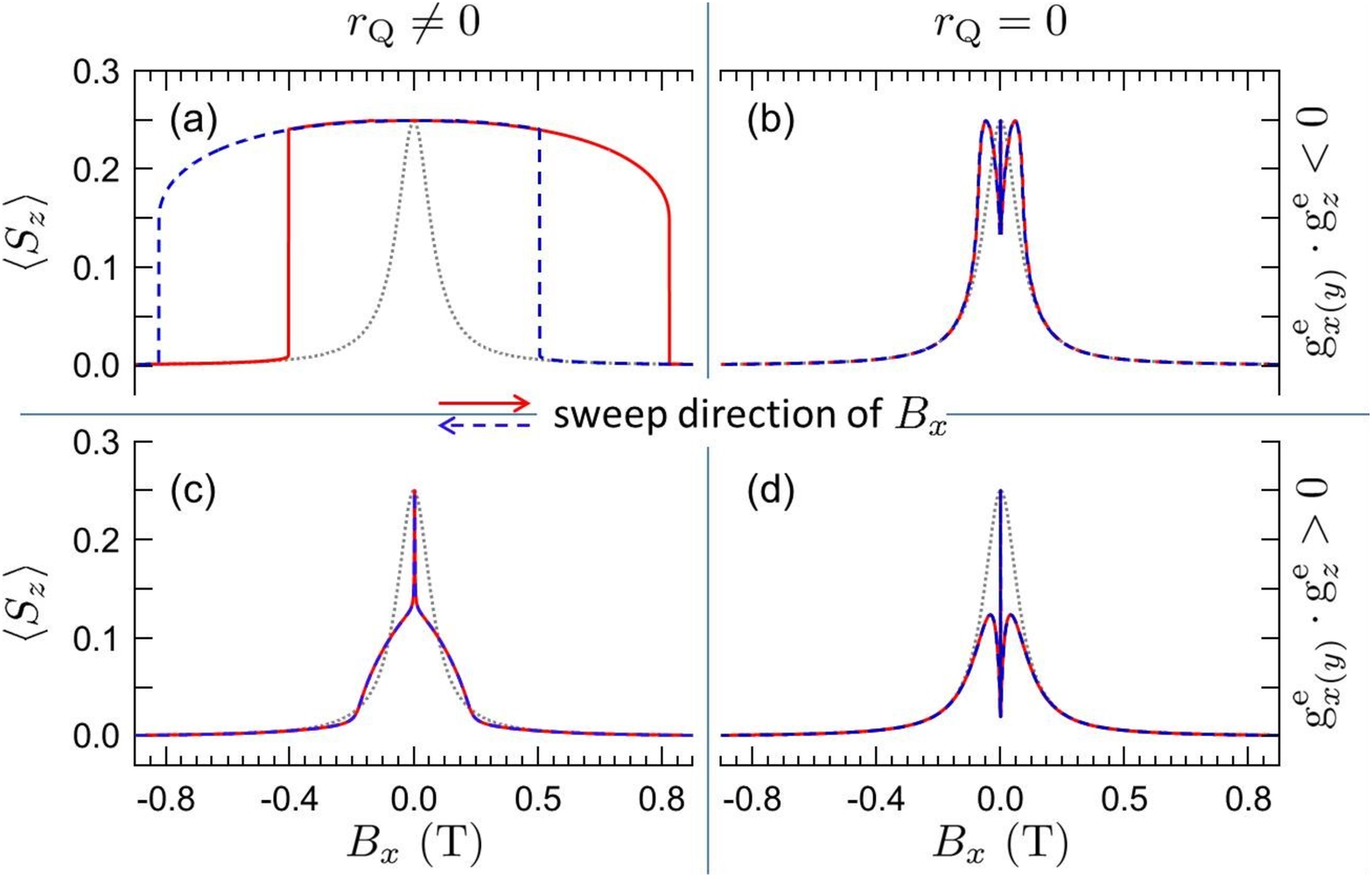}
\end{center}
\caption{(color online) The typical calculated $\langle S_z \rangle$ depending on QI and the sign of $g^{\rm e}_x$. Here $g^{\rm e}_z$ is set to be negative ($g^{\rm e}_z<0$). (a) $r_{\rm Q} \neq 0, \ g^{\rm e}_x >0$, (b) $r_{\rm Q} = 0, \ g^{\rm e}_x >0$, (c) $r_{\rm Q} \neq 0, \ g^{\rm e}_x <0$, (d) $r_{\rm Q} = 0, \ g^{\rm e}_x <0$. $g^{\rm n}_k$ is assumed to be isotropic for $k=x,y,z$. Other parameters are same in all cases. In the case of $r_{\rm Q} \neq 0$, $\bm q$ is set along $z$-axis. The dotted gray line represents a Lorentzian Hanle curve with the typical InAlAs QD parameters.~\cite{comment} }
\label{Calc1}
\end{figure}

Moreover, it is necessary to consider the effective magnetic field experienced by nuclear spins, $\bm{B}_{\rm T}^{\rm{(n)}}$ explicitly. 
%As is the case with $\bm{B}_{\rm T}^{\rm{(e)}}$, the sum of the external field and the Knight field $\bm{B}_{\rm e}$, which is caused by the electron spin polarization and is proportional to $\langle \bm{S} \rangle$, determines the $\bm{B}_{\rm T}^{\rm{(n)}}$ ($\equiv \bm{B}_{\rm{ext}} + \bm{B}_{\rm e}$). 
As is the case with $\bm{B}_{\rm T}^{\rm{(e)}}$, $\bm{B}_{\rm T}^{\rm{(n)}}$ is determined by the sum of the external field and the Knight field $\bm{B}_{\rm e}$, i.e. $\bm{B}_{\rm{ext}} + \bm{B}_{\rm e}$. 
Here, $\bm{B}_{\rm e}$ refers to the effective field caused by the electron spin polarization and is proportional to $\langle \bm{S} \rangle$.
The introduction of $\bm{B}_{\rm e}$ explains the Lorentzian Hanle curve with W-shaped dip~\cite{OptOrientation} which appears in a very low magnetic field region ($B_{\rm{ext}}$$\lesssim$0.1 T) as shown later (Fig.~\ref{Calc1}(b)). 
%Actually, the experimental results observed in AlGaAs bulks~\cite{OptOrientation}, ensembles of SA-InGaAs/GaAs QDs~\cite{Kuznetsova13}, and single droplet GaAs/AlGaAs QDs~\cite{Sallen14} have been reproduced well. 
However, it is not possible to describe the observed anomalous Hanle curves with a quite large width (e.g., $|B_{x}^{\rm c}|$$\sim$0.8 T in Fig.~\ref{AHC}(a) and the results in Ref.\onlinecite{Krebs10}) in this framework. Therefore, we improve the conventional model by introducing the effects of QI.  
%In order to calculate the Hanle curve, which is affected by the all components of NSP, it is necessary to introduce the different values for $1/T_{{\rm NF},k}$. On the other hand, $1/T_{{\rm ND},k}$ is assumed to be isotropic for simplicity.

\begin{figure}[tb]
\begin{center}
\includegraphics[width=240pt]{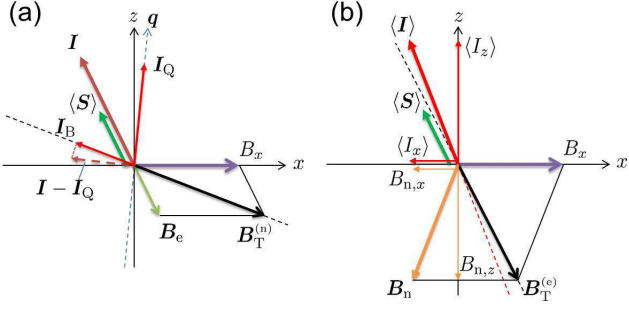}
\end{center}
\caption{(color online) Schematics of the spin polarizations and the resultant effective fields. (a) A portion of the HFI-induced $\bm{I}$ is protected by QI and the other one is preserved by $\bm{B}_{\rm T}^{\rm{(n)}}$: $\bm{I}_{\rm Q}$ and $\bm{I}_{\rm B}$. For generality, the vector $\bm{q}$ is tilted slightly from $z$-axis.   (b) The protected NSP $\langle \bm{I} \rangle$ ($= \bm{I}_{\rm Q} + \bm{I}_{\rm B}$) induces $\bm{B}_{\rm n}$ eventually. Since the sign inversion of g-tensor is assumed, $B_{{\rm n}, x}$ is depicted to be parallel to $\langle I_{x} \rangle$ while $B_{{\rm n}, z}$ is anti-parallel to $\langle I_{z} \rangle$.}
\label{vectors}
\end{figure}

One of the significant improvements is the NSP stabilization due to QI. Note that the HFI-induced NSP, $\bm{I}$ must be collinear with the electron spin polarization according to Eq.~(\ref{eq2}), and it evolves under the effect of $\bm{B}_{\rm T}^{\rm{(n)}}$. As depicted in Fig.~\ref{vectors}(a), we assume that the NSP component along the principle axis of QI is preserved partially with a ratio $r_{\rm Q}\ (0 < r_{\rm Q} \le 1)$. Hence the QI-preserving component $\bm{I}_{\rm Q}$ is given as 
\begin{equation}
\bm{I}_{\rm Q} = r_{\rm Q}(\bm{I}\cdot\bm{q})\bm{q}, \label{eqIQ}
\end{equation}
where $\bm{q}$ is an unit vector along with the principle axis of QI. The other component (i.e., $\bm{I} - \bm{I}_{\rm Q}$) is affected by the effective magnetic field, and its projection to $\bm{B}_{\rm T}^{\rm{(n)}}$ termed as $\bm{I}_{\rm B}$ is also preserved as follows:
\begin{equation}
\bm{I}_{\rm B} = \frac{(\bm{I} -\bm{I}_{\rm Q})\cdot \bm{B}_{\rm T}^{\rm (n)}}{|\bm{B}_{\rm T}^{\rm (n)}|} \frac{\bm{B}_{\rm T}^{\rm (n)}}{|\bm{B}_{\rm T}^{\rm (n)}|}.\label{eqIB}
\end{equation}
%Note that the quadrupolar filed $B_{\rm Q}$ cannot induce the spin precession.
Eventually, only the sum of $\bm{I}_{\rm Q}$ and $\bm{I}_{\rm B}$, $\langle \bm{I} \rangle$ survives and acts as a nuclear field according to the following relation:
\begin{equation}
\bm{B}_{\rm n} = \frac{\tilde{A}}{{\bar{g}_{\rm e}}\mu_{\rm B}}(\bm{I}_{\rm Q} + \bm{I}_{\rm B}). \label{NF}
\end{equation}
It should be noted that the survived $\langle \bm{I} \rangle$ is non-collinear with $\langle \bm{S} \rangle$ and the original $\bm{I}$ as shown in Fig.~\ref{vectors} (b). 

In the following model calculations, we assume that the ratio $r_{\rm Q}$ in Eq.~(\ref{eqIQ}) depends on the applied transverse field as follows:
\begin{eqnarray}
r_{\rm Q}(B_{x}) = \frac{r_0}{1+(B_{x}/B_{\rm Q})^2},\label{eqrQ}
\end{eqnarray}
where $B_{\rm Q}$ is a measure of the QI strength as converted to an effective magnetic field (quadrupolar field). Note that the quadrupolar field $B_{\rm Q}$ cannot induce the spin precession. 
%where $B_{\rm Q}$ is a measure of the effective magnetic field induced by QI, the quadrupolar field. 
The parameter $r_{0}$ is an amplitude of the Lorentzian shape, and the condition $r_{0} =1$ is used in the calculations. 
The efficiency of the NSP stabilization is supposed to decrease with increasing $| B_{x} |$ if the principle axis of QI is almost perpendicular to $B_{x}$. This is because the relative strength of QI to $B_x$ reduces with increasing $| B_{x} |$ and the NSP component perpendicular to $B_x$ is easy to relax. 
In our InAlAs QDs, the magnitude of the quadrupolar field is estimated to be $\sim$280 mT from other experiments~\cite{Matsusaki17}.
%The efficiency of the NSP stabilization is supposed to decrease with increasing $| B_{x} |$ if the principle axis of QI is almost perpendicular to $B_{x}$. This is because the relative strength of QI to $B_x$ reduces with increasing $| B_{x} |$, and the NSP component perpendicular to $B_x$ is easy to relax. In our InAlAs QDs, the magnitude of the quadrupolar field is estimated to be $\sim$280 mT from other experiments~\cite{Matsusaki17}.
%In Fig.~\ref{Calc1}, $r_0=1$ and $B_{\rm Q}$=280 mT~\cite{Matsusaki17} are used and the principal axis $\bm q$ of QI is assumed to be directed to $z$-axis for simplicity.

Further, we assume an anisotropic nature in the nucleus \textit{or} electron g-tensor, that is, the sign of the in-plane g-factor is opposite to that of $z$-component. Thus, in Fig.~\ref{vectors}(b), $B_{{\rm n}, x}$ is depicted to be parallel to $\langle I_{x} \rangle$ while $B_{{\rm n}, z}$ is anti-parallel to $\langle I_{z} \rangle$. 
%This is because the direction of the resultant ${B}_{{\rm n},k}$ is determined by the signs of ${g}^{\rm e}_k$ and ${g}^{\rm n}_k$.
Because ${B}_{{\rm n},k}$ includes the HFI coupling constant $\tilde{A}_k$ and ${g}^{\rm e}_k$, the direction of the resultant ${B}_{{\rm n},k}$ is determined by the signs of ${g}^{\rm n}_k$ and ${g}^{\rm e}_k$. The sign inversion of g-factors in the in-plane and out-of-plane is required to achieve the compensation of the external field $B_{x}$ by in-plane nuclear field $B_{{\rm n}, x}$ that increases by a flip-flop term of HFI.
%; this is another necessary point to reproduce the anomalously broadened Hanle curve and the hysteretic behavior.

Figure~\ref{Calc1} highlights the impacts of the NSP stabilization and the sign inversion of g-factors. %The calculated Hanle curves are the steady state solutions of Eqs.~(\ref{Bloch}) and (\ref{eq2}).  
In this framework, the calculated results can be classified by two important parameters: the sign of $g^{\rm e}_{x(y)} \cdot g^{\rm e}_{z}$ (or $g_{x(y)}^{\rm n} \cdot g_{z}^{\rm n}$) and $r_{\rm Q}$ in Eq.~(\ref{eqIQ}). %Other parameters are same in all of the Hanle curves. 
The latter represents the presence or absence of $\bm{I}_{\rm Q}$.
In Fig.~\ref{Calc1}, we set $g^{\rm e}_z<0$ and the nucleus g-factors $g_{k}^{\rm n}$ are assumed to be isotropic. %Thus the HFI constant $\tilde A_k$ is equivalent for $k=x,y,z$.
In the figure, (a) $r_{\rm Q} \neq 0, \ g^{\rm e}_x >0$, (b) $r_{\rm Q} = 0, \ g^{\rm e}_x >0$, (c) $r_{\rm Q} \neq 0, \ g^{\rm e}_x <0$, (d) $r_{\rm Q} = 0, \ g^{\rm e}_x <0$, and all the other parameters are identical in (a)-(d)~\cite{comment4}. In the calculations, the relaxation time of the NSP $T_{{\rm ND},k}$ is assumed to be isotropic.

%Figure~\ref{Calc1} (a) and (b) indicate the impact of QI-induced NSP stabilization clearly; $r_{\rm Q} \neq 0$ in (a) and $r_{\rm Q}  =0$ in (b).
%Note that the sign of the HFI constants $\tilde A_x$ and $\tilde A_z$ coordinates the sign of $g^{\rm n}_x$ and $g^{\rm n}_z$, respectively, because of $\tilde A_k \propto g^{\rm n}_k $, which affects the direction of $B_{{\rm n},k}$ according to Eq.~(\ref{NF}). 
%In this calculation, the magnitude of the HFI constant $|\tilde A_k|$ is assumed to be equivalent for $k=x,y,z$, but $\tilde A_{x(y)}$ has a different sign of $\tilde A_{z}$.
In the cases of $r_{\rm Q}$=0 ((b) and (d)), the effect of a tilting $\bm B_{\rm n}$ is reproduced as discussed in Ref.~\onlinecite{OptOrientation}. $\bm B_{\rm n}$ cooled by $\bm B_{\rm e}$ inclines by the applied $B_x$. The W-shaped dip (i.e. recovery of DCP) in (b) may correspond to the actual observations in AlGaAs bulks~\cite{OptOrientation} and single droplet GaAs/AlGaAs QDs~\cite{Sallen14}. Since an emerged $B_{{\rm n},x}$ by tilting of $\bm B_{\rm n}$ is small and nearly constant, the compensation (b) or enhancement (d) of $B_x$ by $B_{{\rm n},x}$ occurs within a low $B_x$ region ($\le$0.1 T), and therefore, the depolarization is converging quickly to the tails of a normal Lorentzian curve regardless of a sweep direction of $B_x$.

In the case of (a) ($r_{\rm Q} \neq 0$ and $g^{\rm e}_x \cdot g^{\rm e}_z <0$), the Hanle curves indicates a much larger full width than $2B_{1/2}$. The extension of the width is induced by the compensation of $B_x$ by $B_{{\rm n}, x}$ that follows the change in $B_x$, and therefore, the bistable (hysteretic) behavior appears. 
On the other hand, the effect of $r_{\rm Q} \neq 0$ is strongly reduced in (c) since $B_{{\rm n},x}$ enhances the total effective field and the growth of $B_{{\rm n},x}$ is terminated by high energy cost in the flip-flop process. Therefore, only slight broadening of the curve occurs. 
%On the other hand, although $r_{\rm Q} \neq 0$, there is only slight broadening of the curve in the case of (c) ($g^{\rm e}_x \cdot g^{\rm e}_z >0$) since $B_{{\rm n},x}$ enhances the total effective field and the growth of $B_{{\rm n},x}$ is terminated by high energy cost in the fli-flop process.
%On the other hand, regardless of $r_{\rm Q} \neq 0$, there is no broadening of the curve in the case of (c) ($g^{\rm e}_x \cdot g^{\rm e}_z >0$) since $B_{{\rm n},x}$ enhances the total effective field and the growth of $B_{{\rm n},x}$ is terminated by high energy cost in the flip-flop process.

\begin{figure}[t]
\begin{center}
\includegraphics[width=240pt]{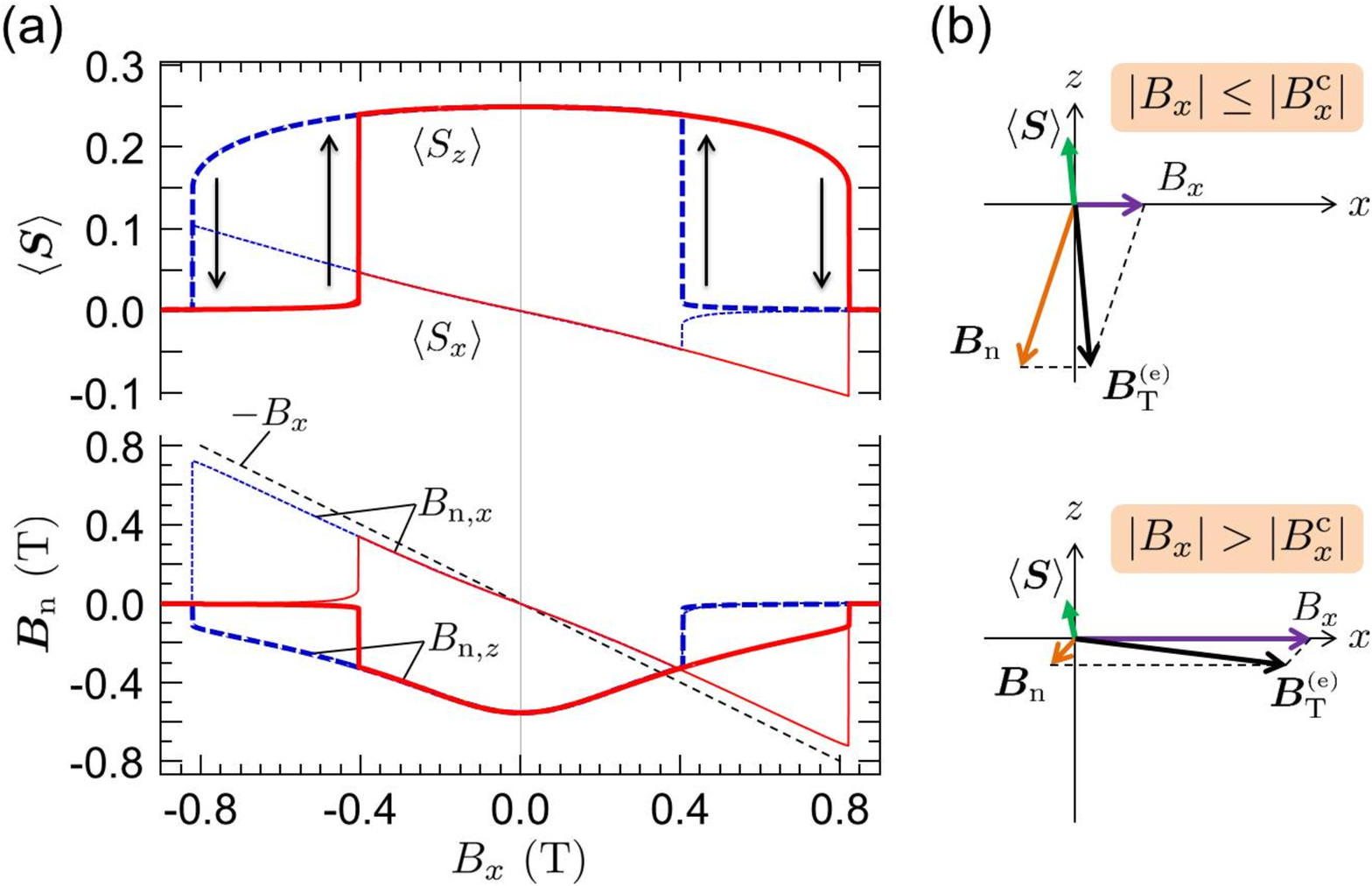}
\end{center}
\caption{(color online) (a) The calculated $\langle S_{z} \rangle$ (thick lines) and $\langle S_{x} \rangle$ (thin lines) in upper panel, and $B_{{\rm n},z}$ (thick lines) and $B_{{\rm n},x}$ (thin lines) in lower panel under the condition, $r_{\rm Q} \neq 0$ and $g_{x}^{\rm n} \cdot g_{z}^{\rm n} <0$. The solid (dashed) lines represent the results with increasing (decreasing) $B_x$. $\langle {S}_y \rangle$ and $B_{{\rm n},y}$ are not shown here for an easy view.~\cite{comment4} (b) Schematics of $\langle \bm{S} \rangle$ and $\bm{B}_{\rm n}$ in $x$-$z$ plane corresponding to the case $|B_{x}| \le |B_{x}^{\rm c}|$ (upper panel) and $|B_{x}| > |B_{x}^{\rm c}|$ (lower panel).}
\label{Calc2}
\end{figure}

\begin{figure}[tbh]
\begin{center}
\includegraphics[width=240pt]{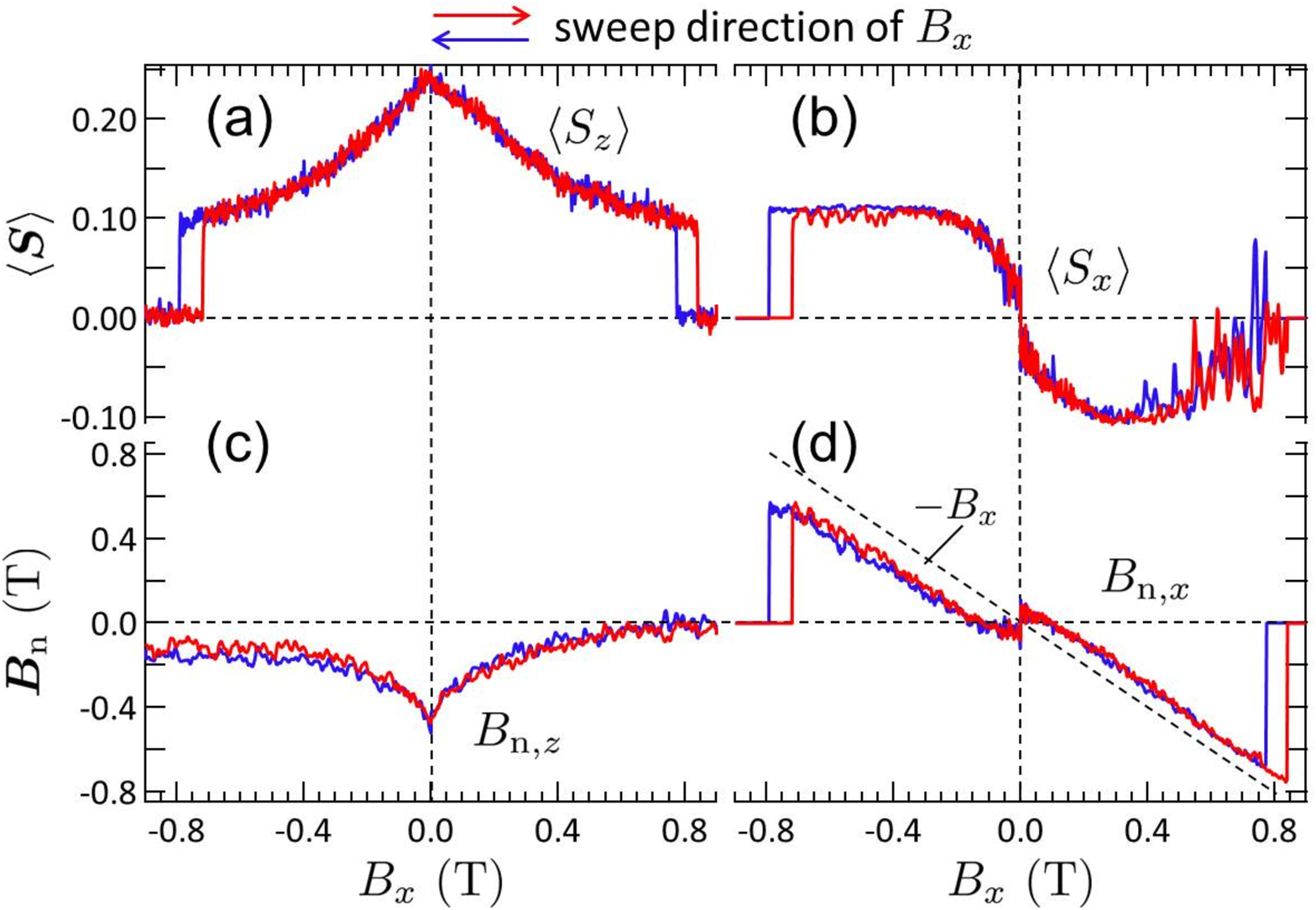}
\end{center}
\caption{(color online) (a) The observed $\langle S_{z} \rangle (= \rho_{\rm c}/2)$, (b) the expected $\langle S_{x} \rangle$, (c) the observed $B_{{\rm n},z}$, and (d) the expected $B_{{\rm n},x}$. Red (blue) line represents the result with increasing (decreasing) $B_x$. (b) and (d) are obtained by substituting the observed $\langle S_{z} \rangle$ and $B_{{\rm n},z}$ to Eqs.~(\ref{sssolux}) and (\ref{sssoluz}).}
\label{Exp2}
\end{figure}
Figure~\ref{Calc2} (a) shows the calculated results about the components of $\langle \bm{S} \rangle$ and $\bm{B}_{\rm n}$ under the condition corresponding to the case of Fig.~\ref{Calc1} (a). However, in Fig.~\ref{Calc2}, the sign inversion is present in nucleus g-factors (i.e. $g_{x}^{\rm n} \cdot g_{z}^{\rm n} <0$) and both conditions of g-factors (Figs.~\ref{Calc1} and~\ref{Calc2}) give the same results basically~\cite{comment3}.  
In the figure, $\langle S_{y} \rangle$ and $B_{{\rm n}, y}$ are not shown for an easy view because of their small magnitudes within $|B_x|<|B_x^{\rm c}|$.
%Figure~\ref{Calc2} (a) shows the calculated results about the components of $\langle \bm{S} \rangle$ and $\bm{B}_{\rm n}$ under the condition corresponding to the case of Fig.~\ref{Calc1} (a). However, in Fig.~\ref{Calc2}, the sign inversion is present in nucleus g-factors (i.e. $g_{x}^{\rm n} \cdot g_{z}^{\rm n} <0$) and both conditions in the sign inversion of g-factors (Figs.~\ref{Calc1} and~\ref{Calc2}) give the same results basically in the case of $r_{\rm Q} \neq 0$~\cite{comment3}. In the figure, $\langle S_{y} \rangle$ and $B_{{\rm n}, y}$ are not shown for an easy view because of their small magnitudes within $|B_x|<|B_x^{\rm c}|$. 

The calculations reproduce well the experimental results in Fig.~\ref{AHC} at the following characteristic points;
%Figure~\ref{Calc2} (a) which corresponds to the case of Fig.~\ref{Calc1}(a) shows the results with $\langle S_x \rangle$ and $B_{\rm n}$ in detail. The calculated results reproduce well the experimental results in Fig.~\ref{Fig2} at the following characteristic points; 
\begin{enumerate}
  \item {a large value of $\langle {S_z} \rangle$ is preserved even under a large $|B_{x}|$ until the critical field $B_{x}^{\rm c}$,}
  \item {$\langle {S_z} \rangle$ changes abruptly at $B_x^{\rm c}$ and shows the hysteretic (thus, bistable) behavior,}
  \item {$\langle {S_z} \rangle$-curve is symmetric with respect to the sweep direction of $B_{x}$,}
  \item {$| B_{{\rm n}, z}|$ reduces gradually with increasing $|B_{x}|$.}
%  \item {a large $\langle {S_z} \rangle \ (\propto \rho_{\rm c}/2$) is preserved till a large $|B_x^{\rm c}|$,}
%  \item {$\langle {S_z} \rangle$ changes suddenly at the critical field $B_x^{\rm c}$ and shows the hysteretic (thus, bistable) behavior,}
%  \item {$\langle {S_z} \rangle$ changes symmetrically with respect to the sign of $B_x$ and sweep direction,}
%  \item {$B_{{\rm n},z}$ reduces gradually with increasing $|B_x|$.}%'±'±'ɍ€–Ú'ð''­
\end{enumerate}
The schematics of the electron spin polarization and the nuclear field are summarized in Fig.~\ref{Calc2} (b); the upper (lower) panel corresponds to the case $|B_{x}| \le |B_{x}^{\rm c}|$ ($|B_{x}| > |B_{x}^{\rm c}|$). In the small-$B_{x}$ region, the in-plane component of $\bm{B}_{\rm n}$ compensates the applied field, and the effective field $\bm{B}_{\rm T}^{\rm{(e)}}$ is almost parallel to $z$-direction. Therefore, a large value of $\langle S_{z} \rangle$ is kept under the condition $|B_{x}| \le |B_{x}^{\rm c}|$.  In the large-$B_{x}$ region, on the other hand, since the induced $\bm{B}_{\rm n}$ is quite small, the vector $\langle \bm{S} \rangle$ feels a large transverse field and $\langle S_{z} \rangle$ relaxes quickly.
%On the other hand, since the induced $\bm{B}_{\rm n}$ is quite small in the large-$B_{x}$ region, the vector $\langle \bm{S} \rangle$ feels a large transverse field and $\langle S_{z} \rangle$ relaxes quickly.
%As a result, as shown in Fig.\ref{Calc2} (b), an electron spin feels only a longitudinal field in the small $B_{x}$ region where a high $\langle S_{z} \rangle$ (i.e. DCP) is kept. On the other hand, as shown in Fig.\ref{Calc2} (c), at $|B_{x}| > |B^{\rm c}_{x}|$, there is almost zero $\bm B_{\rm n}$ so that an electron spin feels the $B_{x}$ directly and $\langle S_{z} \rangle$ relaxes quickly. 

Note that the $B_{x}$-dependence of $r_{\rm Q}$ (as shown in Eq.~(\ref{eqrQ})) is necessary to reproduce the gradual reduction in $B_{{\rm n}, z}$; if $r_{\rm Q}$ is independent of $B_{x}$ and has a constant value, the calculated $B_{{\rm n}, z}$ shows a similar shape with $\langle S_{z} \rangle$ and abrupt changes occur at $|B_{x}^{\rm c}|$~\cite{comment2}. 
Further, a tilting angle of $\bm q$ from the $z$-direction induces the asymmetry and horizontal shift of the Hanle curve.

Finally, we provide the $B_{x}$-dependence of $\langle S_{x} \rangle$ and $B_{{\rm n},x}$ expected from experimental data of $\langle S_{z} \rangle$ and $B_{{\rm n},z}$ in Fig.~\ref{Exp2}. They are obtained from the steady state solutions of the Bloch equation (see Appendix~\ref{APDXB})  by substituting the observed $\langle S_{z} \rangle = \rho_{\rm c}/2$ and $B_{{\rm n},z} = \Delta E_{\rm{OHS}}/({g}^{\rm e}_{z} \mu_{\rm B})$ and solving for $\langle S_{x} \rangle$ and $B_{{\rm n},x}$. Comparing the expected $|\langle S_{x} \rangle |$ (Fig.~\ref{Exp2} (b)) with the computed $|\langle S_{x} \rangle |$ (Fig.~\ref{Calc2} (a)), the expected one has the maximal value and shows saturation or decay while the computed $|\langle S_x \rangle|$ increases monotonically with increasing $|B_x|$. Asymmetry in the expected $|\langle S_x \rangle|$ with respect to $B_x$ is originated from the asymmetry of $B_{{\rm n},z}$, which has small but finite value in a large negative $B_x$ region. 

As a whole, the proposed model can reproduce the observed results qualitatively. However, the quantitative disagreement of the hysteresis width of $|B_x^{\rm c}|$ between the calculated and observed results could not be resolved at present in the range of the parameters we changed.
Despite such a disagreement, $|B^{\rm n}_{x}|$ agrees well with the computed results. These suggest that there is a possibility that another mechanism for the in-plane nuclear field formation may work in a relatively large $|B_{x}|$ region. One plausible candidate is a non-collinear HFI process~\cite{Huang10}, which is not included in our proposed model. Although the main origin of the growth of $I_{x}$ in a proposed model is a flip-flop process between a finite $S_{x} $ and  $I_{x}$, the cooperation of the non-collinear HFI (i.e. $\propto I_{x} S_{z}$) contributes to form $I_{x}$ from $S_{z} $, and may improve the qualitative agreement in relatively large $B_x$ region. 
In addition, the introduction of the $B_x$-dependence into $\tau_{\rm c}$ and $T_{\rm ND}$ may improve the quantitative features.
%In addition, the incorporation of the $B_x$-dependence into $\tau_{\rm c}$ and $T_{\rm ND}$ may improve the quantitative agreement. 

\section{Conclusion}
We investigated the in-plane nuclear field formation via Hanle effect measurements of $X^{+}$ in single self-assembled InAlAs quantum dots. The observed Hanle curves showed anomalously large full width and hysteretic behavior which cannot be explained by an existing model. To reproduce the measured anomalies, we propose a phenomenological model including the nuclear quadrupolar effect and the sign inversion between in-plane and out-of-plane g-factors, which induce the compensation of $B_{x}$ by $B_{{\rm n},x}$. The quadrupolar splitting allows a part of NSP to be preserved nearly along $z$-axis, which induces the $\langle S_{x} \rangle$ that can change $I_x$ via collinear HFI. The proposed model reproduces well the characteristic points of the observed anomalous Hanle curve, and consequently it has the good qualitative agreement. Therefore, we conclude that the collinear HFI is a dominant mechanism for the in-plane nuclear field formation although the non-collinear HFI may contribute to form the in-plane field in relatively large externally applied field.

\acknowledgements
The authors would like to acknowledge H. Sasakura for sample growth and fruitful discussions.
This work is supported by JSPS KAKENHI (Grants No. 26800162 and No. 17K19046) and the Asahi Glass Foundation.

\appendix
\section{in-plane electron g-factors}\label{APDXA}

\begin{figure}[t]
\begin{center}
\includegraphics[width=220pt]{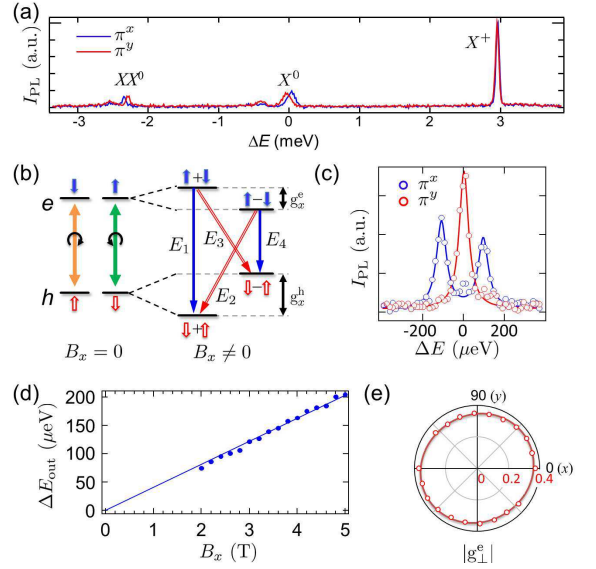}
\end{center}
\caption{(color online) (a) Polarization-resolved PL spectra of a typical SA-InAlAs QD under non-polarized excitation at 6 K and 0 T. The horizontal axis is replotted from the midpoint of the $X^{0}$ doublet. (b) A level diagram of the hole (open arrows) and electron (solid arrows) states under a transverse magnetic field $B_x$. (c) The $X^+$ PL spectra at $B_{x}$=5 T detected in the ($\pi^{x},\ \pi^{y}$) basis. 
The origin of the horizontal axis was set to the energy of the inner PL peaks. (d) The observed Zeeman splitting of the outer peaks as a function of $B_{x}$. The solid line is a fitting curve. (e) Polar plot of the magnitude of in-plane electron g-factor, $|{g}^{\rm e}_{\perp}|$.}
\label{GF}
\end{figure}

Here, we evaluate the in-plane electron g-factor, which is one of key parameters to describe the Hanle curves and coupled electron-nuclear spin dynamics. 
Figure~\ref{GF}(a) shows the polarization-resolved PL spectra under non-polarized excitation at 6 K and 0 T of a single InAlAs QD used in this study. 
The spectra indicate three emissions: the neutral biexciton ($XX^0$), neutral exciton ($X^0$), and positive trion ($X^+$) from the low energy side. Each charge state is assigned by considering the fine structure splitting (FSS) and the binding energy. 
The FSS of $\sim$73 $\mu$eV, the inverse pattern of FSS in the $X^0$ and $XX^0$ peaks, and no splitting in the $X^+$ peak are observed clearly. 

As shown in Fig.~\ref{GF}(b), a transverse magnetic field $B_x$ mixes the spin-up and spin-down states for the electron of $X^+$ and hole, respectively, and the four transitions ($E_{1},\ E_{2},\ E_{3},\ E_{4}$) with linearly polarized emissions and absorptions become optically active. The Zeeman splittings of the outer peaks $\Delta E_{\rm out}=|E_1-E_4|$ with $\pi^{x}$-polarization and of the inner peaks $\Delta E_{\rm in}=|E_2-E_3|$ with $\pi^{y}$-polarization are given by $\Delta E_{\rm out} = (|{g}^{\rm e}_x|+|{g}^{\rm h}_x| )\mu_{\rm B}B_x$ and $\Delta E_{\rm in} = (|{g}^{\rm e}_x|-|{g}^{\rm h}_x| )\mu_{\rm B}B_x $, respectively.
Figure~\ref{GF}(c) shows the polarization-resolved $X^{+}$ PL spectra at 5 T under non-polarized excitation. In this QD, there is no splitting in $\pi^y$-PLs within our spectral resolution, and it indicates that the magnitude of $g^{\rm e}_x$ is very close to ${g}^{\rm h}_x$. Figure~\ref{GF}(d) shows the observed energy splitting of $\pi^{x}$-PLs in a range of 2-5 T. The observed $\Delta E_{\rm{out}}$ increases with $B_{x}$, and from the line fitting, the magnitudes of the electron and hole g-factors are deduced to be $|{g}^{\rm e}_x|$$\simeq |{g}^{\rm h}_x| = 0.35 \pm 0.01$.
Moreover, the anisotropy of the in-plane electron g-factor $|{g}^{\rm e}_{\perp}|$ was investigated by rotating the QD sample around $z$-axis. Fig.~\ref{GF}(e) is a polar plot of $|{g}^{\rm e}_{\perp}|$, and it indicates that the in-plane g-factor anisotropy is negligible. 
In addition to the in-plane g-factors, the electron and hole g-factors in $z$-direction of this QD were evaluated independently to be ${g}^{\rm e}_{z}=-0.34 \pm 0.02$ and ${g}^{\rm h}_{z}=2.57 \pm 0.01$ by the method canceling an optically-induced nuclear magnetic field in $z$-direction by a longitudinal field~\cite{Matsusaki17}. 
The obtained set of the electron g-factor, (${g}_{x}^{\rm e}$, ${g}_{y}^{\rm e}$, ${g}_{z}^{\rm e}$) is used in the section of model calculations.

\section{Bloch equation}\label{APDXB}
The steady state solutions of the Bloch equation (Eq.~(\ref{Bloch})) are written as
\begin{align} 
\langle {\bm{S}} \rangle = 
\frac{S_0}{B_{1/2}^2+(B_x+B_{{\rm n},x})^2
+B_{{\rm n},y}^2
+B_{{\rm n},z}^2
} \nonumber \\
\times
\begin{bmatrix}
B_{1/2}B_{{\rm n},y} + B_{{\rm n},z}(B_x+B_{{\rm n},x} \\
B_{{\rm n},y} B_{{\rm n},z} -
B_{1/2}(B_x+B_{{\rm n},x}) \\
B_{1/2}^2 + B_{{\rm n},z}^2
\end{bmatrix}. 
\end{align}
Omitting the component $B_{{\rm n},y}$ which is very small in the simulated results, the $x$ and $z$ components of $\langle {\bm{S}} \rangle$ can be represented analytically as
\begin{align}
\langle S_x \rangle&=S_0 \frac{(B_x+B_{{\rm n},x})B_{{\rm n},z} }{B_{1/2}^2+(B_x+B_{{\rm n},x})^2+B_{{\rm n},z}^2 },  \label{sssolux}
 \\
\langle S_z \rangle&=S_0 \frac{B_{1/2}^2+B_{{\rm n},z}^2}{B_{1/2}^2+(B_x+B_{{\rm n},x})^2+B_{{\rm n},z}^2 }.   \label{sssoluz}
\end{align}
where $B_{1/2}\ (=\hbar/(|{g}^{\rm e}_{\perp}| \mu_{\rm B} T_{\rm s}))$ is a half width of the normal Lorentzian curve. 
Eqns.~(\ref{sssolux}) and (\ref{sssoluz}) were used to obtain the expected $\langle S_{x} \rangle$ and $B_{{\rm n},x}$ from the observed $\langle S_{x} \rangle$ and $B_{{\rm n},z}$ in Fig.~\ref{Exp2}.


\begin{thebibliography}{00}
\bibitem{OptOrientation} \textit{Optical Orientation}, Modern Problems in Condensed Matter Sciences Vol. 8, Chaps. 2 and 5, edited by F. Meier and B. Zakharchenya (North-Holland, NewYork, 1984).

\bibitem{SpinPhysics} \textit{Spin Physics in Semiconductors}, Springer Series in Solid-State Sciences Vol. 157, Chaps. 1 and 11, edited by M. I. Dyakonov (Springer, Berlin, 2008).

\bibitem{Urbaszek13} Recent optical investigation of nuclear spin physics in QDs are reviewed comprehensively: B. Urbaszek, X. Marie, T. Amand, O. Krebs, P. Voisin, P. Maletinsky, A. H\"{o}gele, A. Imamoglu, Rev. Mod. Phys. \textbf{85}, 79 (2013).
% Nuclear spin physics in quantum dots: An optical investigation

\bibitem{Gammon01} D. Gammon, Al. L. Efros, T. A. Kennedy, M. Rosen, D. S. Katzer, D. Park, S. W. Brown, V. L. Korenev, and I. A. Merkulov, Phys. Rev. Lett. \textbf{86}, 5176 (2001).
% Electron and Nuclear Spin Interactions in the Optical Spectra of Single GaAs Quantum Dots

\bibitem{Yokoi05} T. Yokoi, S. Adachi, H. Sasakura, S. Muto, H. Z. Song, T. Usuki, and S. Hirose, Phys. Rev. B \textbf{71}, 041307(R) (2005).
% Polarization-dependent shift in excitonic Zeeman splitting of self-assembled In0.75Al0.25As?Al0.3Ga0.7As quantum dots


\bibitem{Eble06} B. Eble, O. Krebs, A. Lema\^{i}re, K. Kowalik, A. Kudelski, P. Voisin, B. Urbaszek, X. Marie, and T. Amand, 
Phys. Rev. B \textbf{74}, 081306(R) (2006).
%Dynamic nuclear polarization of a single charge-tunable InAs?GaAs quantum dot



\bibitem{Braun06} P.-F. Braun, B. Urbaszek, T. Amand, X. Marie, O. Krebs, B. Eble, A. Lema\^{i}re, and P. Voisin, 
Phys. Rev. B \textbf{74}, 245306 (2006).
% Bistability of the nuclear polarization created through optical pumping in In1?xGaxAs quantum dots

\bibitem{Tartakovskii07} A. I. Tartakovskii, T. Wright, A. Russell, A. B. Van'kov, J. Skiba-Szymanska, I. Drouzas, R. S. Kolodka, M. S. Skolnick, P. W. Fry, A. Tahraoui, H.-Y. Liu, and M. Hopkinson, Phys. Rev. Lett. \textbf{98}, 026806 (2007).
% Nuclear Spin Switch in Semiconductor Quantum Dots

\bibitem{Maletinsky07} P. Maletinsky, C. W. Lai, A. Badolato, and A. Imamoglu, Phys. Rev. B \textbf{75}, 035409 (2007).
% Nonlinear dynamics of quantum dot nuclear spins

\bibitem{Kaji08} R. Kaji, S. Adachi, H. Sasakura, and S. Muto, Phys. Rev. B \textbf{77}, 115345 (2008).
% Hysteretic response of the electron-nuclear spin system in single In0.75Al0.25As quantum dots: Dependences on excitation power and polarization

\bibitem{Slichter96} C. P. Slichter, \textit{Principles of Magnetic Resonance}, Chapter 10 (Springer, 1996), 3rd ed.

\bibitem{Latta09} C. Latta, A. H\"{o}gele, Y. Zhao, A. N. Vamivakas, P. Maletinsky, M. Kroner, J. Dreiser, I. Carusotto, A. Badolato, D. Schuh, W. Wegscheider, M. Atature, and A. Imamoglu, Nat. Phys. \textbf{5}, 758 (2009).
% Confluence of resonant laser excitation and bidirectional quantum-dot nuclear-spin polarization

\bibitem{Xu09} X. Xu, W. Yao, B. Sun, D. G. Steel, A. S. Bracker, D. Gammon, and L. J. Sham, Nature \textbf{459}, 1105 (2009).
% Optically controlled locking of the nuclear field via coherent dark-state spectroscopy

\bibitem{Hogele12} A. H\"{o}gele, M. Kroner, C. Latta, M. Claassen, I. Carusotto, C. Bulutay, and A. Imamoglu, Phys. Rev. Lett. \textbf{108}, 197403 (2012).
%Dynamic Nuclear Spin Polarization in the Resonant Laser Excitation of an InGaAs Quantum Dot

\bibitem{Krebs10} O. Krebs, P. Maletinsky, T. Amand, B. Urbaszek, A. Lema\^{i}re, P. Voisin, X. Marie, and A. Imamoglu, Phys. Rev. Lett. \textbf{104}, 056603 (2010).
% Anomalous Hanle Effect due to Optically Created Transverse Overhauser Field in Single InAs=GaAs Quantum Dots
%\bibitem{Chekhovich10} E. A. Chekhovich, M. N. Makhonin, J. Skiba-Szymanska, A. B. Krysa, V. D. Kulakovskii, M. S. Skolnick, and A. I. Tartakovskii, Phys. Rev. B \textbf{81}, 245308 (2010).
% Dynamics of optically induced nuclear spin polarization in individual InP/GaxIn1?xP quantum dots

\bibitem{Chekhovich12} E. A. Chekhovich, K. V. Kavokin, J. Puebla, A. B. Krysa, M. Hopkinson, A. D. Andreev, A. M. Sanchez, R. Beanland, M. S. Skolnick and A. I. Tartakovskii, Nat. Nanotech. \textbf{7}, 646 (2012).
% Structural analysis of strained quantum dots using nuclear magnetic resonance
\bibitem{Chekhovich15} E. A. Chekhovich, M. Hopkinson, M. S. Skolnick, and A. I. Tartakovskii, Nat. Commun. \textbf{6}, 6348 (2015).
%Suppression of nuclear spin bath fluctuations in self-assembled quantum dots induced by inhomogeneous strain

%\bibitem{Stockill16} R. Stockill, C. Le Gall, C. Matthiesen, L. Huthmacher, E. Clarke, M. Hugues, and M. Atat\"{u}re, Nat. Commun. \textbf{7}, 12745 (2016).
% Quantum dot spin coherence governed by a strained nuclear environment


%\bibitem{Slichter96} C. P. Slichter, Principles of Magnetic Resonance (Springer, 1996), 3rd ed.

\bibitem{Dzhioev07} R. I. Dzhioev and V. L. Korenev, Phys. Rev. Lett. \textbf{99}, 037401 (2007).
% Stabilization of the Electron-Nuclear Spin Orientation in Quantum Dots by the Nuclear Quadrupole Interaction

\bibitem{Maletinsky08} P. Maletinsky, doctoral thesis (Swiss Federal Institute of Technology, Zurich, 2008).
% Polarization and Manipulation of a Mesoscopic NUclear Spin Ensemble Using a Single Confined Electron Spin


\bibitem{Nilsson13} J. Nilsson, L. Bouet, A. J. Bennett, T. Amand, R. M. Stevenson, I. Farrer, D. A. Ritchine, S. Kunz, X. Marie, A. J. Shields, and B. Urbaszek, Phys. Rev. B \textbf{88}, 085306 (2013).

%\bibitem{Bracker05} A. S. Bracker, E. A. Stinaff, D. Gammon, M. E. Ware, J. G. Tischler, A. Shabaev, Al. L. Efros, D. Park, D. Gershoni, V. L. Korenev, and I. A. Merkulov, Phys. Rev. Lett. \textbf{94}, 047402 (2005).
% Optical Pumping of the Electronic and Nuclear Spin of Single Charge-Tunable Quantum Dots

\bibitem{Sallen14} G. Sallen, S. Kunz, T. Amand, L. Bouet, T. Kuroda, T. Mano, D. Paget, O. Krebs, X. Marie, K. Sakoda, and B. Urbaszek, Nat. Commun. \textbf{5}, 3268 (2014).

\bibitem{Matsusaki17} R. Matsusaki, R. Kaji, S. Yamamoto, H. Sasakura, and S. Adachi, arXiv:1703.06046 (2017).

\bibitem{comment} The width of the normal Hanle curve is determined by the spin lifetime $T_{\rm s}$ and g-factor $g_{x}^{\rm e}$ of a localized electron. Here, $T_{\rm s}$ and $|g_{x}^{\rm e}|$ are assumed to be 0.5 ns and 0.35, respectively, as typical values for single InAlAs QDs.


%\bibitem{Abragam} \textit{The Principle of Nuclear Magnetism}, A. Abragam (Oxford University Press, Oxford, UK, 1961).

%\bibitem{Paget77} D. Paget, G. Lampel, B. Sapoval, and V. I. Safarov, Phys. Rev. B \textbf{15}, 5780 (1977).
% Low field electron-nuclear spin coupling in gallium arsenide under optical pumping conditions

\bibitem{comment4} The following parameters are used in the calculations: 
$f_{\rm e}$=0.015, $\tau_{\rm c}$=100 ps, $T_{{\rm ND},k}$=10 ms, $B_{\rm Q}$=280 mT, $r_0$=1.0, and $\bm q$ is assumed to be along $z$-axis. In Fig.~\ref{Calc2}, $g^{\rm e}_{x(y,z)}$=-0.35, $\tilde A_{x(y)}$=-52.6 $\mu$eV, $ \tilde A_{z}$=+52.6 $\mu$eV, $g^{\rm n}_{x(y)}$=-0.179, $g^{\rm n}_{z}$=+0.179, and $g^{\rm e}_k (k=x,y,z)$=-0.35 are assumed.

%\bibitem{Kuznetsova13} M. S. Kuznetsova, K. Flisinski, I. Ya. Gerlovin, I. V. Ignatiev, K. V. Kavokin, S. Yu. Verbin, D. R. Yakovlev, D. Reuter, A. D. Wieck, and M. Bayer, Phys. Rev. B \textbf{87}, 235320 (2013).
%Hanle effect in (In,Ga)As quantum dots: Role of nuclear spin fluctuations



\bibitem{comment3} The out-of-plane g-factor ${g}^{\rm e}_{z}$ of InAlAs QDs has an opposite sign of that of InAs QDs~\cite{Kaji14}. The sign of ${g}^{\rm e}_{x(y)}$ may be the same as that of ${g}^{\rm e}_{z}$ because of the isotropic nature of a conduction electron. 
In contrast, the nucleus g-factor is considered to have a large anisotropy because of QI. Considering that the anomalous Hanle curves have been observed in both kinds of SA-QDs and those facts, it is likely that the requirement of the sign inversion in the in-plane and out-of-plane of g-factors of this study could be attributed to the nucleus g-factors.

\bibitem{Kaji14} R. Kaji, T. Hozumi, Y. Hachiyama, T. Tomii, H. Sasakura, M. Jo, and S. Adachi, Appl. Phys. Express \textbf{7}, 065002 (2014). 
%"Dispersions of hole and electron g-factors in single InAs quantum rings evaluated using optically induced nuclear spin polarization",

\bibitem{comment2} As shown in the lower panel of Fig.~\ref{Calc2}(a), small discontinuities at $|B_x^{\rm c}|$ arise in the model calculations, while the experimental results change continuously (Fig.~\ref{Exp2}(c)). We guess tentatively that the step in Overhauser shift might be so small that we failed to detect it. 

\bibitem{Huang10} C.-W. Huang and X. Hu, Phys. Rev. B \textbf{81}, 205304 (2010).
% Theoretical study of nuclear spin polarization and depolarization in self-assembled quantum dots





%\bibitem{Krebs07} O. Krebs, B. Eble, A. Lema\^{i}re, B. Urbaszek, K. Kowalik, A. Kudelski, X. Marie, T. Amand, and P. Voisin, Phys. Stat. Solidi. A \textbf{204}, 202 (2007).
% Role of hyperfine interaction on electron spin optical orientation in charge-controlled InAs?GaAs single quantum dots

%\bibitem{Krebs08} O. Krebs, B. Eble, A. Lema\^{i}re, P. Voisin, B. Urbaszek, T. Amand, and X. Marie, C. R. Phys. \textbf{9}, 874 (2008).
% Hyperfine interaction in InAs/GaAs self-assembled quantum dots: dynamical nuclear polarization versus spin relaxation

%\bibitem{Bechtold15} A. Bechtold, D. Ranuch, F. Li, T. Simmet, P.-L. Ardelt, A. Regler, K. Muller, N. A. Sinitsyn, and J. J. Finley, Nat. Phys. \textbf{11},1005 (2015).

%\bibitem{Pal07} B. Pal, S. Y. Verbin, I. V. Ignatiev, M. Ikezawa, and Y. Masumoto, Phys. Rev. B \textbf{75}, 125322 (2007).
% Nuclear-spin effects in singly negatively charged InP quantum dots

%\bibitem{Kaji12} R. Kaji, S. Adachi, H. Sasakura, and S. Muto, Phys. Rev. B \textbf{85}, 155315 (2012). 
% Direct observation of nuclear field fluctuations in single quantum dots 



%\bibitem{Adachi12} S. Adachi, R. Kaji, S. Furukawa, Y. Yokoyama, and S. Muto, J. Appl. Phys. \textbf{111}, 103531 (2012).
%"Nuclear spin depolarization via slow spin diffusion in single InAlAs quantum dots observed by using erase-pump-probe technique",

%\bibitem{comment1} The DCP measurement in Section III does not need this improvement because a large $B_{\rm z}$ is applied and the orthogonal PL components that indicate a large Zeeman splitting can be taken in single CCD detection.

%bibitem{Bulutay12} C. Bulutay, Phys. Rev. B \textbf{85}, 115313 (2012).
% Quadrupolar spectra of nuclear spins in strained InxGa1-xAs quantum dots

%\bibitem{Maletinsky09} P. Maletinsky, M. Kroner, and A. Imamoglu, Nat. Phys. \textbf{5}, 407 (2009).
% Breakdown of the nuclear-spin-temperature approach in quantum-dot demagnetization experiments
\end{thebibliography}
\end{document}